\documentstyle[12pt,epsf]{article}
\newcommand{\be}{\begin{equation}}
\newcommand{\ee}{\end{equation}}
\newcommand{\bea}{\begin{eqnarray}}
\newcommand{\eea}{\end{eqnarray}}
\newcommand{\ci}{\cite}
\newcommand{\bi}{\bibitem}
\newcommand{\nono}{\nonumber \\}
\newcommand{\tr}{{\rm tr}}
\newcommand{\p}{{\bf p}}
\newcommand{\e}{{\rm e}}
\newcommand{\R}{{\bf R}}

\newcommand{\ssf}{{\sin^2 F}}

\newcommand{\da}{\dagger}
\newcommand{\dd}{\partial}
\newcommand{\bftau}{\mbox{\boldmath$\tau$}}
\newcommand{\bfo}{\mbox{\boldmath$\omega$}}

\newcommand{\half}{\frac{1}{2}}

\def\dal{\,\lower0.3ex\vbox{\hrule\hbox{\vrule\kern2pt\vbox{\kern4pt\kern4pt}
\kern2pt\vrule}\hrule}\,}

\def\s{\sigma}
\def\o{\omega}

\begin{document}
\title{{\sl The nucleus as a fluid of skyrmions: Energy levels and
nucleon properties in the medium}\thanks{dedicated to
the memory of Prof. Judah M. Eisenberg}}
\vspace{1 true cm}
\author{G. K\"albermann\\
Faculty of Agriculture and Racah Institute of Physics\\
Hebrew University, Rehovot 76100, Israel}
\maketitle

\begin{abstract}
\baselineskip 1.5 pc
A model of a fluid of skyrmions coupled to a scalar and to the
$\o $ meson mean fields is developed.
The central and spin-orbit potentials of a skyrmion
generated by the fields predict correct energy levels
in selected closed shell nuclei.
The effect of the meson fields on the properties of
skyrmions in nuclei is investigated.
\end{abstract}

\newpage
\baselineskip 1.5 pc

\section{\label{Introduction} Introduction: Fluid models of nuclei}

Fluid models of nuclei have been used for quite a long time.
The time-honored liquid drop model\ci{text} is extremely
successful in describing the ground state bulk properties of nuclei by
utilizing extremely simple assumptions. These assumptions are
based on the observation of the saturation phenomenon
in nuclei as the predominant effect in determining the
binding energies.
The analogy of the nucleus to a fluid is also borne out
in the almost constant density of nuclei throughout the
periodic table and the lack of localization of
nucleons around fixed positions, as it would be for a solid,
due to zero point motion. The nucleus is a quantum fluid
and cannot be modeled as a usual fluid for which the molecules behave
to a large extent classically.

More recently, the need to include relativistic effects
in the description of hadron dynamics inside the nucleus
lead to the development of the nuclear mean field models\ci{ser1}.
These theories can be viewed as fluid models in which the
mean field plays the role of the continuum (hydrodynamical regime),
while the elementary fermion fields are their microscopic source.
Again, the model has proven fruitful in the prediction
of ground states as well as collective states of nuclei.

On the other hand, nuclei -especially light ones- possess
shell structures that are usually traced back to the single particle
energy levels.
This is clearly seen in the oscillations of
nuclear binding energies as a function of mass, as well as in
the existence or inexistence of stable nuclides.

Heavy nuclei, however, seem to behave as deformed
rigid bodies.
This collective behavior is put in evidence
by the large quadrupole excitation amplitudes.

The question of the nature of the state of matter in the
nucleus consequently depends on both the excitation energy,
the mass and the density of the nucleus.

In any event, the nucleus cannot be compared
to a solid. Even heavy nuclei have a rather soft incompressibility
factor. Being compressible, they resemble more a gas or
a compressible liquid, as advocated by the propugnators
of the liquid drop model.

If one agrees with this view of a nucleus as a fluid, liquid
or gas depending on the temperature, then it is valid to consider its
description in terms of a few degrees of freedom as it is done for conventional
fluids. The quantum nature of the nucleons demands the treatment to be
quantal. The high speeds attainable by nucleons in the nucleus
requires a relativistic treatment.

One key ingredient that is absent in all the hydrodynamical models
is the nucleon finite size.
The charge radius of the proton is not negligible compared to nuclear
dimensions, in contradistinction to molecules in a liquid.
In the nonrelativistic Hartree-Fock or relativistic
mean field  Hartree treatments of nuclei, the nucleon is treated  as
pointlike. Some information about the
finite extent of the nucleon is sometimes included
in the form of short range correlations, short range hard-core, etc.

The purpose of the present work is to try to deal in a
coherent manner with the fluid aspect of nuclei and the finite
size of the nucleon.

Much has been learned in the past 40 years about the nucleon.
Despite the fact that QCD is the accepted theory of strong
interactions, it is practically impossible to use it as a tool
to generate baryons as confined objects of quarks and gluons,
and even less to describe nuclei. We here resort to a
low energy model of strong interactions, namely, the Skyrme model\ci{skyrme}.
The issue of finite nucleon size is dealt with by describing
the nucleon as a topological soliton, the skyrmion.
The fluid aspect is addressed by using a relativistic
mean field theory in the spirit of the Walecka model\ci{ser1}.
The scheme is simplified by resorting to a dilute fluid
approximation to be explained below. In this
manner the nucleons are essentially free,
interactions being mediated by the mean fields.
Section 2 summarizes the formalism and previous results
both for nuclear matter and finite nuclei.
Section 3 is devoted to the energy levels of a skyrmion in the nucleus.
Section 4 treats the effects of the meson fields on the
properties of the skyrmion.

\section{\label{fluid}A skyrmion fluid model of nuclei}

In the early 60's Skyrme developed a topological soliton model
for nucleons\ci{skyrme}. In this model,
baryons emerge as classical topological
solutions of a nonlinear meson lagrangian. More than
a decade ago this perspective was taken up by Witten\ci{witten}. He showed
that baryons may indeed appear as solitons in the large $N_c$ limit of QCD.
The development of collective coordinate quantization of
soliton rotations \ci{anw} gave a tremendous boost to the field.
The door was now open for the investigation of baryons and nuclei.
The Skyrme model and its topological solitons: skyrmions, may be regarded
as the effective degrees of freedom of low energy baryon physics.
This is especially
appealing due to the difficulty of solving QCD explicitly.

The Skyrme model has had moderate success in dealing with the
nucleon-nucleon interaction, including the attractive isoscalar central
potential for which several mechanisms were proposed\ci{ek1995}.
It is difficult to solve the model exactly for
large baryon number structures.
Static solutions of various
geometries for some special baryon number
cases were recently found using the rational map ansatz\ci{houghton}.
These solutions are of a limited practical use
for the treatment of nuclei. Nuclei appear to be extremely dynamical
assembly of nucleons. The special static solutions
found in the literature resemble a crystal\ci{piette}.
Some of these solutions are
in conflict with the spin-isospin assignment of nuclear ground states\ci{irwin}.
We here opt for an approach in which the nucleus is
viewed as a dynamical aggregate of nucleons.
For this purpose it is convenient to
follow the path of fluid-like mean field theories, such as the
relativistic model of Walecka\ci{ser1}.

In the relativistic mean field theory of Walecka the nucleons are described
using Dirac wave functions. The nucleons interact with
meson fields. The meson fields are in turn determined by the
various densities of the baryons(scalar, vector,etc.).
The nucleons have no internal structure. They are considered
elementary pointlike objects.
The model is simple enough to handle the basic features of
nuclei and nuclear matter, and yet powerful enough to
predict ground states and excited states with a relatively
small amount of parameters that enter in the meson
self-interactions and meson-nucleon coupling constants.
The model is used at the first quantized level.
Quantum corrections spoil the predictions.
Being only a low energy effective model for nuclei, it is
valid to consider it within its realm of applicability:
the mean field level.

The same path will be followed here.
We will construct and use a mean field model of
nucleons in terms of skyrmions, instead of Dirac pointlike objects.
The task becomes more complicated due to
the finite volume occupied by the skyrmion. In some sense the
Dirac wave function represents only the center of mass
dynamics, while, the first quantized skyrmion will
carry information about the interior of the nucleon.

The skyrmion lagrangian is built by demanding
both isospin and chiral symmetries (SU(2)xSU(2)).
The many-body model developed here still obeys the same
principles.

The mean fields considered in the present work are the minimal set
in order to be able to describe low energy nuclear phenomenology.
A scalar field provides the attractive interaction and binds
nuclei.
The coupling of the scalar to the soliton is taken
from phenomelogically succesful models\ci{schech,ru1,ellis}.
The added scalar brings about the intermediate range attraction
between skyrmions\ci{yabu}, that is absent for
potentials generated from the product ansatz.

In order to stabilize the nucleus and reproduce the saturation
property we also need a repulsive interaction. We here
follow the Walecka models and use a meson field that is an isoscalar-vector
field, the $\o$ meson.

Consider a field theory lagrangian of skyrmions, a scalar $\psi$ and the
$\o$ meson  \ci{ek1995}
\bea \label{skydil}
 {\cal L} & = & {\cal L}_{2\ {\psi}} +
 {\cal L}_2 + {\cal L}_4 - V_{\rm interaction}
 - V(\psi) + {\cal L}_\o\nono
& = & \half\dd_\mu\psi\,\dd^\mu\psi
-  g_{\psi}\frac{\psi}{\gamma}{F_\pi^2 \over 16}\tr(L_\mu L^\mu)
+ {1 \over 32 e^2}\tr[L_\mu,\, L_\nu]^2 -  g_V~~ \o_\mu B^\mu\nono
&-& V(\psi) - {1\over 4} {(\dd_\mu\o_
\nu - \dd_\nu\o_\mu)^2} + \frac{\psi^2}{2\gamma^2} m_\o^2~~\o_\mu^2.
\eea
Here
\be \label{Lmu}
L_\mu \equiv U^\da\dd_\mu U,
\ee
where $U({\bf r},t)$ is the chiral field, $F_\pi$ is the pion decay constant
and $e$ the Skyrme parameter, $\psi$ is the scalar,
$\gamma$, an energy scale-parameter, and, $g_{\psi}$, a coupling constant.
We have  omitted the $\rho$ meson as it
is our intention to treat symmetric nuclei only.

In the present work we will not specify the potential
of the scalar. It must contain a mass-term and higher order terms.

Due to the fact that the sign of $\psi$ is irrelevant,
it leads to an attractive interaction for both positive and negative signs,
we can redefine the field in the form $\psi=\gamma~e^{\s}$.
We will henceforth absorb the coupling constant $g_{\psi}$ in the
definition of the constant $F_{\pi}$, that is in any event a
fitted parameter in the Skyrme model.

Similarly to the mean field theory of pointlike baryons, we consider
an ensemble of essentially free skyrmions. Each skyrmion will be
accompanied by its own scalar field and $\o$ fields. Although we deal with
free skyrmions, the average properties of the ensemble are still
included in distribution functions that depend on the
density and temperature.

The skyrmion is  a topological soliton.
The winding number of the soliton was already identified
by Skyrme with the baryon charge. It is absolutely
conserved regardless of the dynamics. It is a geometrical property,
not a Noether charge.
We have to take care of this aspect when building many body
ansatze for the nucleon fluid. The baryon number should be
exactly conserved.

An ansatz that
conserves baryon number exactly and still allows for a reasonable
treatment
of the dynamics of $N$ skyrmions is the product ansatz

\be \label{pa}
U_{B=N}({\bf r},\R_1,\R_2,\cdots,\R_N) = U({\bf r}-\R_1)\,U({\bf r}-\R_2)\cdots
U({\bf r}-\R_N),
\ee

Where $R_i$ is the location of the center of skyrmion $i$.

This ansatz is not an exact solution of the skyrmion sector of the problem.
We mentioned static exact solutions \ci{houghton} as being
appropriate for a crystal-like structure of nuclei.
It is our intention to treat the dynamics of the nucleons in the nucleus
in a reasonable manner without sacrificing nucleon motion, hence we
opt for this crude approximation. Despite being a rough ansatz,
it is very maleable and quite appropriate for the nuclear case in which
the nucleons are not tightly packed.
Nucleons in the nucleus
are separated by an average distance in the order of 2 fm.
The solitons overlap at the 'surface' of each
skyrmion. This interaction is replaced by the effective meson
fields. Instead of dealing with the complicated situation of
the N-body interaction in all its aspects, a
small set of mesonic degrees of freedom is chosen
and the skyrmions interact solely with them.
It is expected that the introduction of the mesons in the mean
fields will compensate, at a phenomenological level,
the lack of accuracy induced by the rudimentary product ansatz.

For the meson fields we use  (see section 4 for an improvement
upon this approximation)

\bea \label{additivity}
\s_{B=N} & = & \s_1 + \s_2 + \cdots+ \s_N, \nono
& = &\s_0 + \delta\s_1 + \delta\s_2 +\cdots+ \delta\s_N, \\
\o_{B=N} & = & \o_1 + \o_2 + \cdots+ \o_N, \nono
& = &\o_0 + \delta\o_1 + \delta\o_2 +\cdots+ \delta\o_N,
\eea

Where $\s_0 , \o_0$ are the mean field constant values of the
scalar and the $\o$ and $\delta\s, \delta\o$ represent the
fluctuations.
In thermal equilibrium, the mean field fields will depend on the
temperature $T$ and the chemical potential $\mu$.
Self-consistency then demands
the values of $\s_0$ and $\o_0$ to be determined by the properties
of the ensemble.
In equilibrium the fluctuations will be taken to vanish
in accordance with the mean field approximation.

The topological baryon density

\be \label{Bmu}
B^\mu = \frac{\epsilon^{\mu\alpha\beta\gamma}}{24\pi^2}
\tr \left[\left(U^\da\dd_\alpha U\right)
\left(U^\da\dd_\beta U\right)
\left(U^\da\dd_\gamma U\right)\right],
\ee
with the the product ansatz of eq.~(\ref{pa}) gives

\bea \label{b_1}
B_0 & = & b_1 + b_2 + \cdots + b_N,\nono
\eea

The dynamics of the nucleons is that of a rigid soliton. No internal
excitations are considered. Also skyrmion distortions
will not be allowed, except for swelling and shrinking
of the soliton in the medium.
Only the bulk parameters of the skyrmion,
mass and moment of inertia will be allowed to vary
due to the presence of the mesonic medium.

This is the basis for the dilute fluid approximation.
It is related to the mean
field Dirac model with two major modifications: 1) there is a different
dynamics for the baryons dictated by the lagrangian of eq.~(\ref{skydil}) and,
2) there is a clear way to uncover the baryon response to the medium
through the soliton equation of motion.

The many body skyrmion fluid in the dilute approximation is built from
the product ansatz of eq.~(\ref{pa}) for the skyrmions and the additive
ansatz for the mesons of eq.~(\ref{additivity}).
In the following we use as input
a distribution function for the baryons as a shortcut.

A better method would be one that uses a fully self-consistent
method, such as a Hartree-Fock. However, the use of a definite
distribution function permits a first pass to the problem.
We will use a Thomas-Fermi approximation: The skyrmion fluid
is forced to remain within the boundaries of the nucleus. This
introduces a cutoff Fermi radius. Beyond this radius the
meson fields will still exist, but the baryon densities will
be taken as vanishing.

We stress again that the results obtained here are to be taken
within the context of the two approximations they
hinge upon, namely, a simplistic product ansatz -dilute
fluid approximation- and an averaging
procedure with a pre-chosen distribution function.
The former is a working hypothesis that permits the
treatment of dynamics in a rather straightforward manner,
and, the latter finds support in the the fact that we
want to calculate equilibrium properties.
The first approximation breaks down at short distances between
nucleons and the second loses meaning for light nuclei, for
which the 'mean-fields' are not a wise choice of
degrees of freedom, and direct nucleon-nucleon interactions
are a better way to deal with the problem instead of by means of intermediaries.

The skyrmions dynamics is obtained by boosting
the static skyrmions rigidly. We perform a Lorentz boost
on the collective coordinate $R(t)$ of each skyrmion,
although a non-relativistic Galilean transformation may suffice
and perhaps be more consistent with the spirit of the rigid boosting.
However, using a Lorentz transformation will help us
to find the spin-orbit interaction below.

For the sake
of simplicity consider a Lorentz boost along the $x$ axis with
velocity parameter $v$
(For a different view of the boost problem see ref\ci{ji}):

\bea\label{boost}
x\rightarrow \tilde x & = & \frac {x- R(t)}{\sqrt{1-v^2}} \nono
\tilde y & = & y \nono
\tilde z & = & z \nono
F(\vec{\bf r}) & \rightarrow & F(\vec{\tilde {\bf r}})\nono
\s(\vec{\bf r}) & \rightarrow & \s(\vec{\tilde{\bf r}}) \nono
\o^\mu & \rightarrow & \bigg( \frac{\o(\vec{\tilde{\bf
r}})}{\sqrt{1 - v^2}},\frac {v \o(\vec{\tilde{\bf r}})}{\sqrt{1 -
v^2}}, 0 , 0\bigg) \nono
B^\mu & \rightarrow & \bigg( \frac{B_0(\vec{\tilde{\bf
r}})}{\sqrt{1 - v^2}},\frac {v B_0(\vec{\tilde {\bf r}})}{\sqrt{1
- v^2}}, 0, 0\bigg)
\eea
with $\o^\mu B_\mu= \o_0(\vec{\tilde{\bf r}}) B_0(\vec{\tilde{\bf r}})$.

Introducing the above transformation in eq.~(\ref{skydil}) and
calculating the Hamiltonian, we find the energy of a skyrmion in motion
to be
\be\label{skenergy}
E_p =\bigg(E_2 + E_\s - E_\o\bigg)\frac{2 p^2+ 3 M^2}{3 \epsilon~M} +
E_4~\frac{4~p^2 + 3 M^2}{3 \epsilon~M} +\frac {M}{\epsilon}\bigg(U_\s - U_\o\
+ U_{int}\bigg)
\ee
where
$$\epsilon = \sqrt{p^2 + M^2}, \quad p=\frac{M v}{\sqrt{1 - v^2}}~,$$
$M$ is the skyrmion static mass for nonvanishing $\o_0, \s_0$

\bea \label {mass}
 M(R) & = & 4\pi\int_0^\infty r^2\,dr M(r)\nono
M(r) &= &\e^{2 \s}\frac{F_\pi^2}{8}
\left[F'^2 + 2 \frac{\ssf}{r^2}\right] + \frac{1}{2 e^2}
\frac{\ssf}{r^2} \bigg[\frac{\ssf}{r^2}+2
F'^2\bigg]
\eea
and
\bea\label{en1}
E_2 & = & \frac{4 \pi F_\pi^2}{8} {\int r^2 dr}~e^{2\s}
\bigg( F'^2 + \frac{2 \ssf}{r^2}\bigg) \nono
E_4 & = & \frac{4 \pi}{2 e^2} {\int dr}~\bigg( 2 F'^2 +
\frac{\ssf}{r^2} \bigg) \nono
E_\s & = & \frac{4 \pi \Gamma_0^2}{2} {\int r^2 dr}~e^{2\s}
\s'^2 \nono
E_\o & = & 4 \pi {\int r^2 dr}~\o'^2\nono
U_\s & = & 4\pi {\int r^2 dr}~V_\s\nono
U_\o & = & 4\pi {\int r^2 dr}~e^{2\s}\frac{m_\o^2
\o^2}{2}\nono
U_{int} & = & \frac{2 g_V}{\pi} {\int dr}~\o~F'~\ssf
\eea

If the fields are kept as constant throughout the nucleon volume, we
can use the skyrmion profile equation (or the virial theorem)
to obtain
\be \label{mascale}
M(R) = e^{\s} M_0
\ee
where $ M_0$ is the skyrmion mass for $\s$ = 0. We can include the rotational
energy of the skyrmion in the adiabatic approximation. The result
will yield the same scaling law as above.
An attempt to include
the rotational energy beyond the adiabatic approximation will induce
the known instabilities to radiation of pions.
Henceforth we use the measured free nucleon mass for $M_0$,
that contains both static and rotational energies.
The above scaling is the only effect of the mean fields on the skyrmion, while
the skyrmion contributes to the mean fields through its energy and baryon number
separately. The topological character of the baryon current in the Skyrme model
prevents it from changing. The only effect of the mean fields on the
skyrmion is that of mass scaling.

We can now write down the energy of N skyrmions
in the mean field approximation with a Thomas-Fermi distribution
\be\label{distrib}
f({\p_i},{\R_i})=\Theta(p_F(R_i)-p_i)
\ee
with the normalization condition
\be\label{norm}
\int{d{\R_i}~d{{\p}_i} f}~=1
\ee
The Fermi momentum $p_F$ depends on the distance $R_i$ of the center of
the $i^{th}$ skyrmion from the center of the nucleus. Also, for
indistinguishable particles $p_F(R_i)=p_F(R)$.
The distribution
function for the ensemble is just a product of functions of the type
above.

Averaging over the coordinates $R$ and the corresponding momenta with
the distribution function for zero temperature, the energy of the
nucleus with spherical symmetry we obtain, using eq.(\ref{skenergy})

\bea \label{energy}
E & = & 4 {\pi} \int{ R^2 dR~~E(R)}\nono
E(R) & = & E_{\s} + E_{\o}  +  E_{int} + E_{sk}
\eea
where
\bea \label{detail}
E_{\s} & = & \half~\Gamma_0^2~e^{2\s}\s'^2 + V_{\s}  \nono
E_{\o} & = & -\half~\o'^2-e^{2\s}\frac{m_\o^2\o^2}{2}\nono
E_{int} & = &  g_V~\o~B \nono
E_{sk} & = & \frac {1}{\pi^2} \int_0^{k_P} k^2 dk \sqrt{k^2 + M^2}+
\frac {1}{\pi^2} \int_0^{k_N} k^2 dk \sqrt{k^2 + M^2}
\eea
where $k_P, k_N$ are the proton and neutron local (R dependent)
Fermi momenta.

In eq.~(\ref{detail}) primes denote derivatives with
respect to $R$, whereas in eq.~(\ref{mass}) they represent derivatives with
respect to $r$.
Also
\bea \label {density}
B(R) = \frac {(k_P)^3}{3 \pi^2} + \frac {(k_N)^3}{3 \pi^2}\nono
\eea

The Euler-Lagrange equations for the mean fields become

\bea \label {equations}
{\Gamma}_0^2 ~~ e^{2\s} \bigg(\s'' + 2\s'^2  +  \frac{2\s'}{R} \bigg)  -
\frac{dV_{\s}}{d\s} +m_{\o}^2~\o^2~ e^{2\s} - \frac{\dd E_{sk}}{\dd\s}= 0 \nono
\o'' + \frac {2 \o'}{R}  - m_{\o}^2~\o~e^{2 \s} + g_V B = 0 \nono
\eea
The ground state of the nucleus for fixed number of protons (P) and neutrons
(N) is obtained by minimizing the energy, constrained by Lagrange
multipliers, with respect to the local wavenumbers $k_P$ and $k_N$

\be \label {constraint}
\delta E - {\mu}_P~~\delta P - {\mu}_N~~\delta N~=0
\ee
The algebraic equations for the multipliers become \ci{ser1}

\bea \label {mu}
\mu_P & = & g_V~\o + \half~g_{\rho} b + q~a + \sqrt{k_P^2 + M^2}\nono
\mu_N & = & g_V~\o -\half~g_{\rho} b + \sqrt{k_N^2 + M^2}
\eea

In the Thomas-Fermi approximation, the nucleus has a finite radius
beyond which the nucleon densities are taken to vanish, therefore
${\mu}_P = {\mu}_N$.

In previous works \ci{kal1,kal2}, we found that it is possible to
fit the properties of nuclear matter, including the incompressibility,
 around 270 MeV, and the ground state densities of some magic nuclei,
by introducing a parametrization of the scalar potential
with 4  parameters.

The key restriction for the terms in
the extended potential comes from the requirement that
the expectation value of the $\s$ field in the vacuum remains unchanged
$<vac|\s|vac> = 0$.
Terms of the form $ e^{n\s} - 1$ , generically referred to as no-log terms
\ci{ru1}, are acceptable.

In the present work we will not commit ourselves as to the nature of the
scalar field and its mass. We will not determine the scalar potential
either. The field will be determined phenomenologically from the
nuclear densities.

\section{\label{energy levels}Energy levels of a skyrmion in the nucleus}

In the present work we address the question of the energy levels
and properties of the nucleon in the skyrmion fluid.
To this end we simplify the treatment of section 2.
We solve the $\o$ meson equation~(\ref{equations}) with $B(R)$
replaced by the measured densities and the scalar determined by the
chemical potential. In this manner we avoid tedious parametrizations of
the scalar potential in order to fit the densities. The
assumption is that such a potential exists, and the results of ref.~\ci{kal1}
support it.
The equation of the $\s$ meson is therefore not solved.
The $\o$ meson eq.~(\ref{equations}) is solved starting at the center of
the nucleus up to a cutoff radius $R_c$ at which $\o$=0.
The $\s$ field is then extrapolated beyond that point using
suitable boundary conditions\ci{ser1,kal2}.

A key element of the nucleon interaction in the nucleus is the
spin-orbit potential.
We will here proceed to consider the contribution of both
the scalar and vector fields to this force.
The dominant contribution will arise from the modification
of the mean fields in a rotating frame as viewed by the moving
nucleon. The treatment will be a mixture of relativistic (when essential) and
nonrelativistic methods. A more coherent treatment would call
for a parametrization of the metric of a rotating nucleus, but this
is outside the scope of the present work.\\

Consider first a rigid boost of the skyrmion center
with velocity $\bf v$ of eq.~(\ref{boost})
together with a rigid rotation of
the skyrmion with collective coordinates decribing spin and isospin\ci{witten}

\bea \label {collective}
U({\bf{\tilde r}})~\rightarrow~ A(t) U({\bf{\tilde r}}) A^{\da}(t)
\eea

The spin-orbit interaction may then arise from the
well-known Thomas precession. It can be obtained from two consecutive Lorentz
transformations.
This spin-orbit force will exist regardless of the
extended nature of the skyrmion provided it moves in a central force field
and it carries an {\it{axis}}.
An easy way to implement this transformation is to take the isospin
vector matrix as time dependent

\bea \label{Thomas}
 {\dot{\bftau}} = -{\bf \Omega}_T~\bf x~\bftau
\eea
where ${\bf \Omega}_T$ is the Thomas frequency.
Inserting the above ansatz 
in the skyrmion lagrangian of eq.~(\ref{skydil}) the spin-orbit
interaction at the lowest order in the velocity is found to be
\be\label{so1}
U_{s.o.} \approx-\frac{{\bf S}\cdot{\bf L}}{2~M_0^2~R} \frac{\dd V_C}{\dd R}
\ee
where $V_C$ is the central potential of the skyrmion in the nucleus, $\bf S$
is the spin and $\bf L$ the angular momentum. The same result as found in
standard textbook derivations\ci{jackson}.
In eq.(\ref{so1}) we have used the projection formula \ci{witten}

\bea\label{proj}
{\dot A^{\da}}~A = \frac{-i~{\bftau}\cdot{\bf S}}{2~\lambda(R)}
\eea
where $\lambda(R)$ is the moment of inertia of the nucleon\ci{witten}
\bea\label{lambda}
\lambda(R) & = & \frac{2 \pi}{3}\int~r^2~dr~\Lambda(R)\nono
\Lambda(R) & = & sin^2(F)\bigg[F_{\pi}^2~e^{2\s}~+~\frac{4}{e^2}
\bigg(F'^2+\frac{sin^2(F)}{r^2}\bigg)\bigg]
\eea
and F is the skyrmion profile whose $R$ dependence
enters through the scalar field $\s$.
It turns out, as expected, that the spin-orbit of eq.~(\ref{so1})
is quite negligible, due to the $\frac{1}{M_0^2}$ dependence.

There is another source of spin-orbit interaction. It is due to the
transformation of the fields to a rotating frame, essentially the
coupling of the baryon current to the $\o$ field in a rotating
nucleus analogous to the isoscalar coupling to the photon.
However, the $\o$ meson coupling to the skyrmion, is
not a gauge invariant one and we do not expect the
same result as for the magnetic moment of
the nucleon. The introduction of a gauge invariant coupling
to the skyrmion requires additional terms.

In order to find the spin-orbit potential, we need the mean fields
for a streaming nucleus. In this case there arises a spatial component of the
$\o$ meson field. An appropriate approximate
ansatz for this  component is \ci{ser1}
\bea\label{om}
{\bfo} = {\bf V}~\o_1(R) = ({\bf \Omega~x~R})~\o_1(R)
\eea
where $\bf V$ is the tangential velocity of the nucleus at each $R$ and
$\bf\Omega$
the angular velocity. At the same time the nucleon baryon density develops
a time dependent piece of the form \ci{riska}
\bea\label{b}
{\bf B}({\bf u}) = {\bf S~x~u}~~\frac{B_0(u)}{2~\lambda}
\eea

where $B_0(u)$ is the static baryon density of the skyrmion.

Inserting eqs.(\ref{om},\ref{b}) above  in the lagrangian
(\ref{skydil}), there will appear new terms in energy.
After averaging over the angular directions
of $\R$, we find the equation of motion for $\o_1$ to be
\bea\label{omeq}
\o_1'' + \frac {4 \o_1'}{R}  - m_{\o}^2~\o_1~e^{2 \s} + g_V~B = 0 \nono
\eea
This is very similar to the equation of motion of the static $\o$ in
eq.~(\ref{equations}).
We solve equation (\ref{omeq}) for each nucleus using the
scalar field of the static case and demanding a vanishing
$\o_1$ at infinity.

In order to find the corresponding spin-orbit interaction we consider
a nucleon spinning at rest with a nucleus rotating with a velocity,
${-\bf V}(R)$, opposite to the direction of rotation of the nucleon. Using
the collective coordinate quantization scheme of eq.~(\ref{collective}),
and the projection formula of eq.~(\ref{proj}) we find
\bea\label{uso2}
W_{s.o.} = \frac{-{\bf S} \cdot {\bf L}}{2~M_0~\lambda(R)}~\o_1(R)
\eea

Clearly, $W_{s.o.}$ is more important than
$U_{s.o.}$ due to the $\frac{1}{M_0}$ dependence.
It is a pure skyrmion spin-orbit as evidenced by the presence
of the moment of inertia in the potential.

In the Dirac type of
Walecka models \ci{ser1}, the spin-orbit interaction arises from
the coupling of the lower components of the Dirac wave function.
In conventional nuclear forces calculations
the spin-orbit interaction is found by including pion exchange\ci{pip}.
Here, it arises from the interaction of the rigid rotation of the nucleon
with the flow of the mean fields, which is quite a different mechanism.

The static fields contribute to the energy of the nucleus
an amount proportional to the baryons rest energy, while the
dynamical $\o_1$ contribution originates from the rotational energy.
The latter then adds a small correction to the static mean fields
and may be ignored.

We now focus on the single skyrmion in the nucleus.
Using eq.(\ref{boost}) in the relativistically invariant lagrangian
of eq.~(\ref{skydil}), the classical
Hamiltonian of a single skyrmion in motion becomes

\be\label {ham}
H = \sqrt{p^2 + M^2}+ g_V~\o + W_{s.o.} + U_{s.o.}
\ee
where $p$ is the nucleon momentum, the conjugated variable to the skyrmion
center location $R$.
Expanding the square root in eq.~(\ref{ham}) to order $p^2$
and quantizing the coordinate $R$, we obtain an effective Schr\"odinger equation
for the radial wave function of the skyrmion center with total energy $E$

\bea\label{schroe}
\bigg[\frac{\dd^2}{{\dd R}^2}+\frac{2~\dd}{R~\dd R}
-\frac{l~(l+1)}{R^2}-Q(R)\bigg]{\Psi}= 0\nono
\eea
where
\bea\label{q}
Q(R)  =  e^{2\s}~M_0^2-{\bigg(-g_V~\o +~E~ -W_{s.o.}\bigg )}^2-Z(R)\nono
Z(R)~\approx\frac{g_V}{2 M(R)}\bigg[\frac{{\dd}^2}{{{\dd}R}^2}
+\frac{2~\dd}{R~\dd R}\bigg]\o+ 2~M_0~U_{s.o.}
\eea
The central potential entering the calculation of
$U_{s.o.}$ of eq.(~\ref{so1}) is given by
\bea\label{vc}
V_C = \frac{Q_1(R)}{2 M_0}
\eea
where $Q_1$ is given by $Q(R)$ of eq.(\ref{q}), but without the spin-orbit
pieces. The energy eigenvalue of the Schr\"odinger equation $E$ appears
inside the operator $Q(R)$.

We have solved the Schr\"odinger equation (\ref{schroe}) for the ground state
single particle levels for the magic nuclei
$C^{12} , O^{16}$ and $Ca^{40}$.

The only free parameters are the values of the
scalar and the $\o$ meson at the center of the nucleus.
These parameters determine the chemical potential and the cutoff radius.
They are fixed once for all the energy levels in the chosen nucleus.
We used the measured $\o$ meson and nucleon masses and $g_V$=7.31\ci{kal2}.\\\\
Table 1 shows the comparison between the predicted binding energies and
the experimental ones averaged over proton and neutron states\ci{fuchs,landaud}.

\begin{table}\label{table 1}
\caption {\sl{ Binding energies of single particle levels}}
	\begin{center}
         \medskip
         \begin{tabular}{|c|c|c|c|}
           \hline
Nucleus&Shell&calculated energy&experimental energy\\
&&MeV&MeV\\
&&&\\
           \hline
&&&\\
$C^{12}$&$1s_{\frac{1}{2}}$&36.3&35.2\\
&1$p_{\frac{3}{2}}$&15.7&16.9\\
&&&\\
           \hline
&&&\\
$O^{16}$&$1s_{\frac{1}{2}}$&37.1&43$\pm$5\\
&$1p_{\frac{3}{2}}$&20.5&20.1\\
&$1p_{\frac{1}{2}}$&15.6&13.9\\
&&&\\
           \hline
&&&\\
$Ca^{40}$&$1s_{\frac{1}{2}}$&48&50$\pm$10\\
&$1p_{\frac{3}{2}}$&35&34$\pm$6\\
&$1p_{\frac{1}{2}}$&30.7&34$\pm$6\\
&$1d_{\frac{5}{2}}$&21&18.5\\
&$2s_{\frac{1}{2}}$&15.7&14.5\\
&$1d_{\frac{3}{2}}$&14.8&12\\
&&&\\
           \hline
         \end{tabular}
     \end{center}
\end{table}
\newpage

The results show that the skyrmion picture of both the central
and the spin-orbit interaction is quite good.
The spin-orbit originates solely from the $\o$ meson as viewed by the
rotating skyrmion, in contradistinction to the Dirac mean field case in which
both scalar and vector fields act together to produce a large
interaction.
It might appear at first that
the skyrmion viewed as based upon a $\frac{1}{N_c}$ expansion of QCD,
cannot lead reliably to such fine detalis such as binding energies
of around 20 MeV. However,
the binding energies we find are not clearly related to the $\frac{1}{N_c}$
expansion.
The  $\frac{1}{N_c}$ approximation refers to the properties of the soliton
constructed from colored quarks.
Finite $N_c$ corrections, as well as quantum effects may indeed have influence
on the basic properties of the soliton.
However, the interaction of the skyrmion with the
external scalar and vector fields operating through
location of its center, promoted to the rank of
quantum collective coordinate,
is expected to be relatively unaffected; provided we use the
appropriate parameters for the nucleon, namely its mass and moment of
inertia.

In the Skyrme model, 
the nucleon interacts as a colorless object with its neighbors.
The 'exchanged' meson fields are also colorless.
Whether nucleons are partially deconfined inside the
nucleus and quarks can percolate
and be exchanged between nucleons, remains an open question. 
Nevertheless, in the 
language of skyrmeons, the leading contribution to
the nucleon-nucleon interaction is indeed due
to processes that ignore the color structure of the
baryons, it is color-blind.
This assumptions are based on the fact that the interactions
between nucleons play a role in nuclear structure, only for
internucleon separations that are too large for color forces to
be noticed. The effect of those forces is introduced
phenomenologically in the parameters of the model, whose
scale is not calculated.
In some sense, the skyrmion interactions are like Van der Waals
forces that act between neutral atoms and molecules. Although their
origin resides within electromagnetic interactions, once the
effective forces are determined one can work with them.
The approximation is even better for quarks because they are confined.

\section{\label{modification}Nucleon properties in the medium}

In the previous sections we took the value of the meson fields as
constant throughout the skyrmion. This method yields the
scaling law of eq.~(\ref{mascale}).
In this section we check the validity of this approximation and look for
influences of the skyrmion interior on the
meson fields and viceversa.
Consider now the $\s$ and $\o$ fields radially symmetric, but
allowed to change inside the skyrmion, namely

\bea \label {var}
\s=\s(z)\nono
\o_0=\o_0(z)
\eea

where $z$ is the distance from the center of the nucleus.
Inside a skyrmion we have

\bea\label{z}
z=\sqrt{R^2+r^2-2Rr~cos(\theta_R-\theta_r)}
\eea

with $R$ the center of mass location of the skyrmion and r the radial
distance from this center.

We will simplify the calculation by averaging the fields over angles.
The distribution function we use is independent of angle (eq.~(\ref{distrib})).
The soliton profile will still be taken as spherically
symmetric. We could allow for distortions of the nucleon,
but this will make the calculation extremely cumbersome.
We want to keep the description as simple as possible and still
capture the esence of the effects of the meson fields on the skyrmion.

We average over angles, without determining the meson
fields yet

\bea\label{average}
Q(r,R)=\frac{1}{(4{\pi})^2}\int~d{\Omega}_R\int~d{\Omega}_r\chi(z)\nono
P(r,R)=\frac{1}{(4{\pi})^2}\int~d{\Omega}_R\int~d{\Omega}_r~\o(z)\nono
\eea

With $\chi=e^{2\s}$
obtaining

\bea\label{simple}
Q(r,R)=\frac{1}{4~rR}\int_{|R-r|}^{R+r}{\chi(u)}~ u~ du
\eea

with a similar equation for $P$.

The single skyrmion static energy is now

\be\label{stenergy}
E=4\pi~\int~r^2~dr~\big(Q(r,R)~M(r)-P(r,R)g_V\frac{\ssf~F'}{2{\pi}^2~r^2}\big)
\ee
where $M(r)$ is defined in eq.~(\ref{mass}).
The meson fields depend in a nontrivial way on R and the skyrmion
inherits this dependence.

The skyrmion equation of motion becomes
\bea\label{euler}
\bigg(Q + \frac{8~\ssf}{\tilde r^2}\bigg) F''
& + & F' (2~Q/r +Q') + \frac{4 \sin 2F
F'^2}{\tilde r^2} -\frac{Q \sin 2F }{r^2} \nono
& - & \frac{4 \ssf \sin 2F}{r^2 \tilde r^2} -
\frac{2~g_V~P'~\ssf}{\pi^2~r^2} = 0 \nono
\eea
where $\tilde r = e F_\pi r$ and primes denote derivatives with
respect to r.

We start the calculation by evaluating the
meson fields for the chosen nuclei. The values of the meson
fields at the center of the nucleus are varied in order
to fit the nuclear densities. We do not parametrize
the scalar potential in a definite manner. We take the measured
densities as input and solve the Thomas-Fermi equations of motion
for the $\o$ meson with a fixed chemical potential $\mu$.
The meson fields are then fitted with analytical expressions
that reproduce accurately the calculated values.
For the nuclei $C^{12}$ and $O^{16}$ we use formulae
that are easily integrable in order to
find $Q$ and $P$ of eq.~(\ref{average})

\begin{figure}[tb]
\epsffile{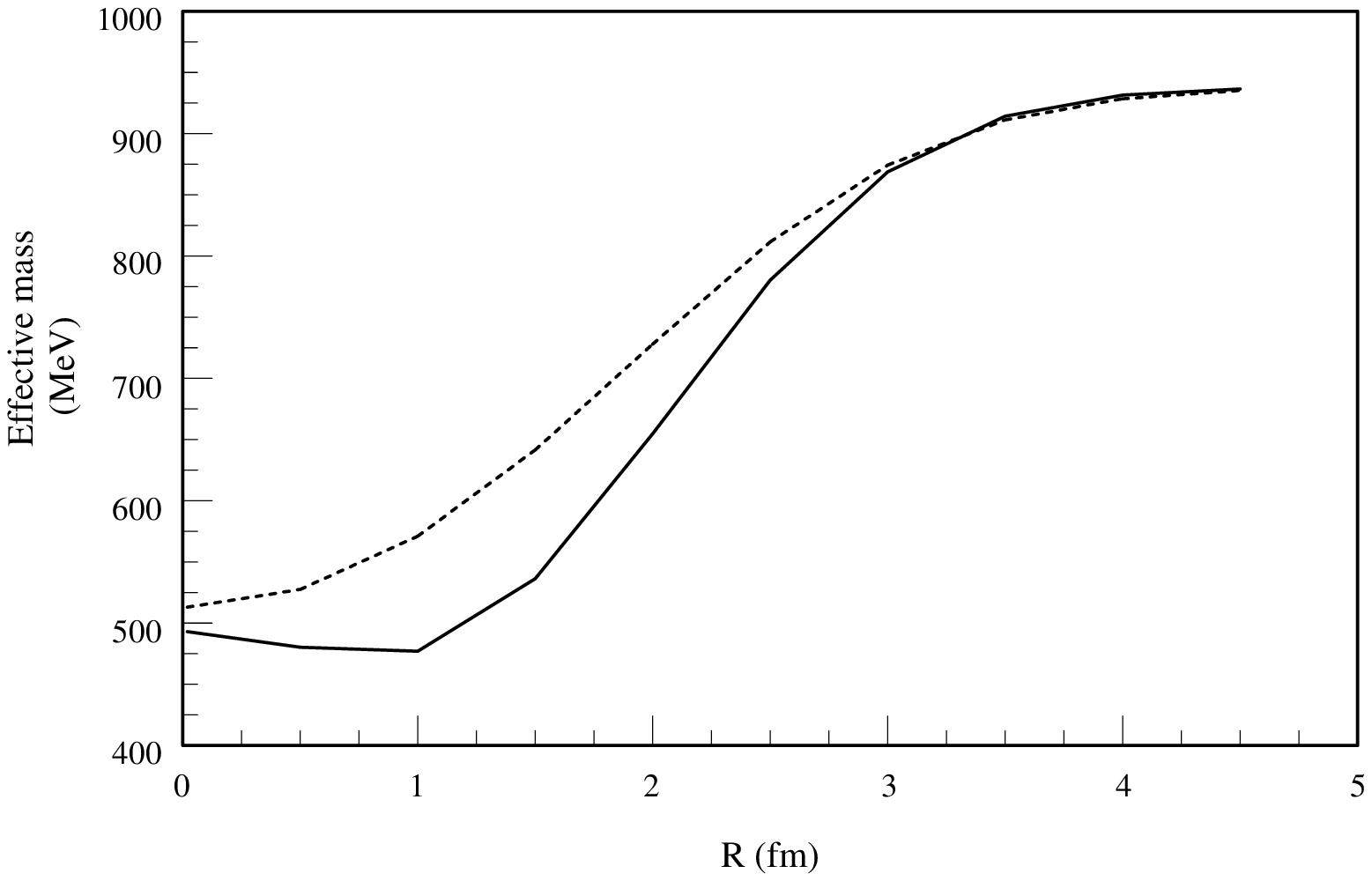}
\vsize=5 cm
\caption{\sl{ $M_{eff}$ in $C^{12}$, constant (modified) meson fields
inside the skyrmion, full line (dashed line)}}
\label{fig1}
\end{figure}
\begin{figure}[tb]
\epsffile{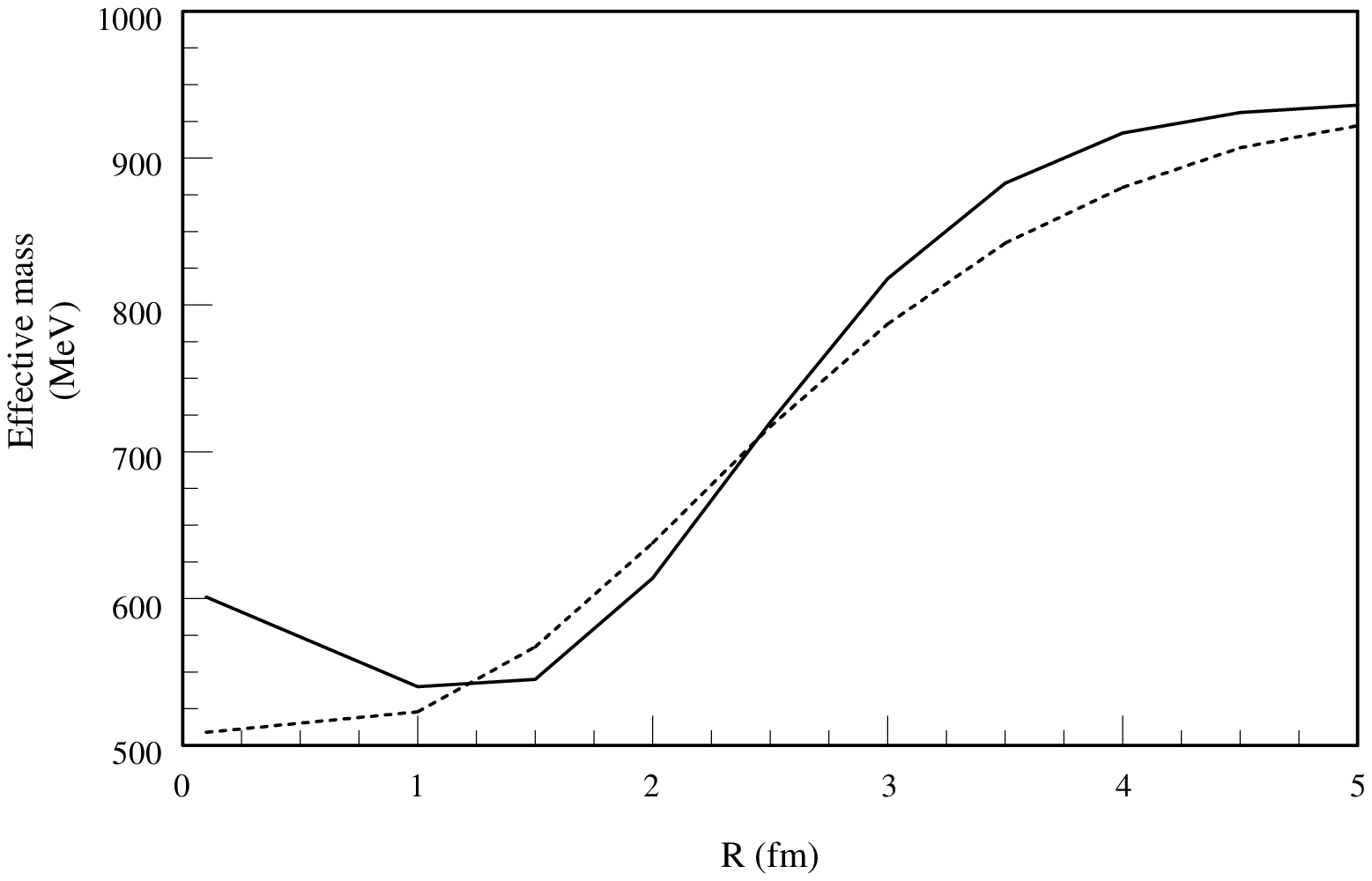}
\vsize=5 cm
\caption{\sl{ $M_{eff}$ in $O^{16}$, constant (modified) meson fields
inside the skyrmion, full line (dashed line)}}
\label{fig2}
\end{figure}
\begin{figure}[tb]
\epsffile{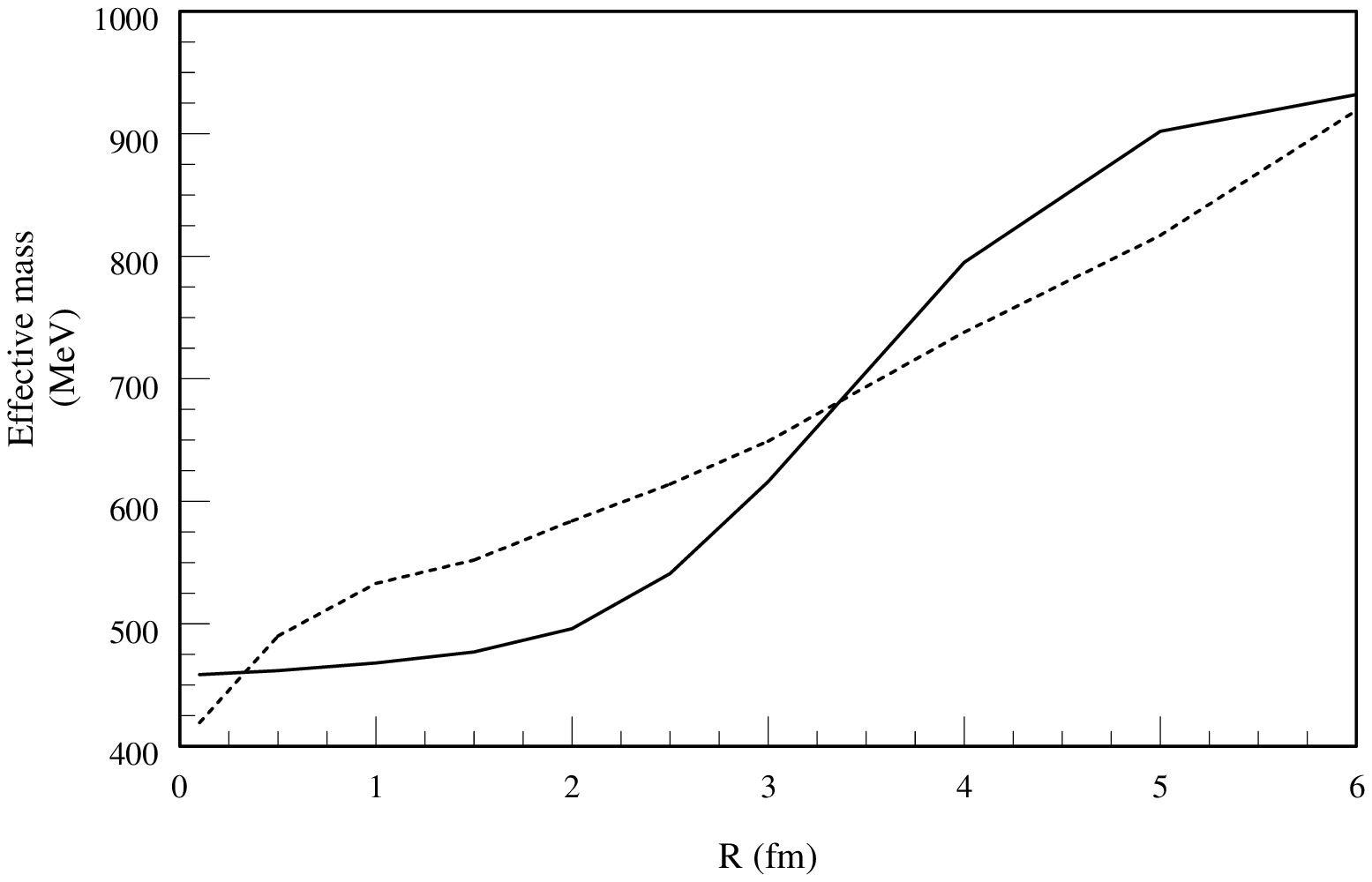}
\vsize=5 cm
\caption{\sl{ $M_{eff}$ in $Ca^{40}$, constant (modified) meson fields
inside the skyrmion, full line (dashed line)}}
\label{fig3}
\end{figure}

\bea\label{fit}
\chi(z)=1+(a_1+b_1~z^2+c_1~z^4)~e^{-d_1~z^2}\nono
\o(z)=(a_2+b_2~z^2+c_2~z^4)~e^{-d_2~z^2}
\eea

whereas for, $Ca^{40}$, it was found after trial and error that
a better suited expression for the $\o$ field is

\bea\label{calcium}
\o(z)=(a_2+b_2~z^2+c_2~z^4)~e^{-d_2~z}
\eea
\begin{table}\label{table2}
\caption {\sl{ Parameters for the fits to the meson fields
with the analytical expressions of eqs.[42,43]}}
	\begin{center}
         \medskip
         \begin{tabular}{|c|c|c|c|}
           \hline
parameter&$C^{12}$&$O^{16}$&$Ca^{40}$\\
\hline
&&&\\
$a_1$& -0.4744275&-0.3581017&-0.5112333\\
$b_1$& -0.2732692&-0.2322270&-0.1049903\\
$c_1$& -7.1264720E-03&-2.4487255E-03&-1.7629359E-02\\
$d_1$&  0.4292937&0.3366868&0.2366882\\
$a_2$&  0.1471073&0.1004175&0.1679764\\
$b_2$&  5.3953879E-02&5.9962720E-02&0.2383440\\
$c_2$&  0.1121303&2.7225710E-02& -9.0451920E-03\\
$d_2$&  0.8262799&0.5333008&0.9995874\\
           \hline
         \end{tabular}
     \end{center}
\end{table}

We then calculate the new soliton profiles using eq.(\ref{euler}),
evaluate the skyrmion effective mass using eq.~(\ref{stenergy})
and compare it to the desired effective mass needed in order
to fit the single particle levels obtained in the previous section
as good as possible.
We iterate the procedure until the
effective masses are the closest to the desired values.

Table 2 shows the parameters for the
analytical expressions of eqs.~(\ref{fit},\ref{calcium}).

Figures 1 through 3 compare the obtained effective skyrmion masses
both with and without meson fields variation inside the skyrmion.

The effective mass of the skyrmion is defined
as $M_0~e^{\s_0}$ for the case of a constant scalar field
inside the nucleon, and

\bea\label{nucmass}
M_{eff}=E+\frac{3}{8 \lambda}
\eea
where E is the nucleon energy obtained from eq.~(\ref{stenergy})
and $\lambda$ is defined in eq.~(\ref{lambda}).
The trend and absolute values of both types of effective masses are
consistent.
However, they are not identical. It was
impossible to improve the agreement beyond the one shown in the
graphs. The inclusion of the volume of the skyrmion in
the evaluation of the meson fields cannot be disregarded.
This is more clearly seen in
figure 4, where we show the scalar field ${\chi}=e^{2\s}$
in the nucleus with and
without the influence of the volume of the skyrmion for $O^{16}$.
A similar picture emerges for the $\o$ field and in the
other nuclei.
\begin{figure}[tb]
\epsffile{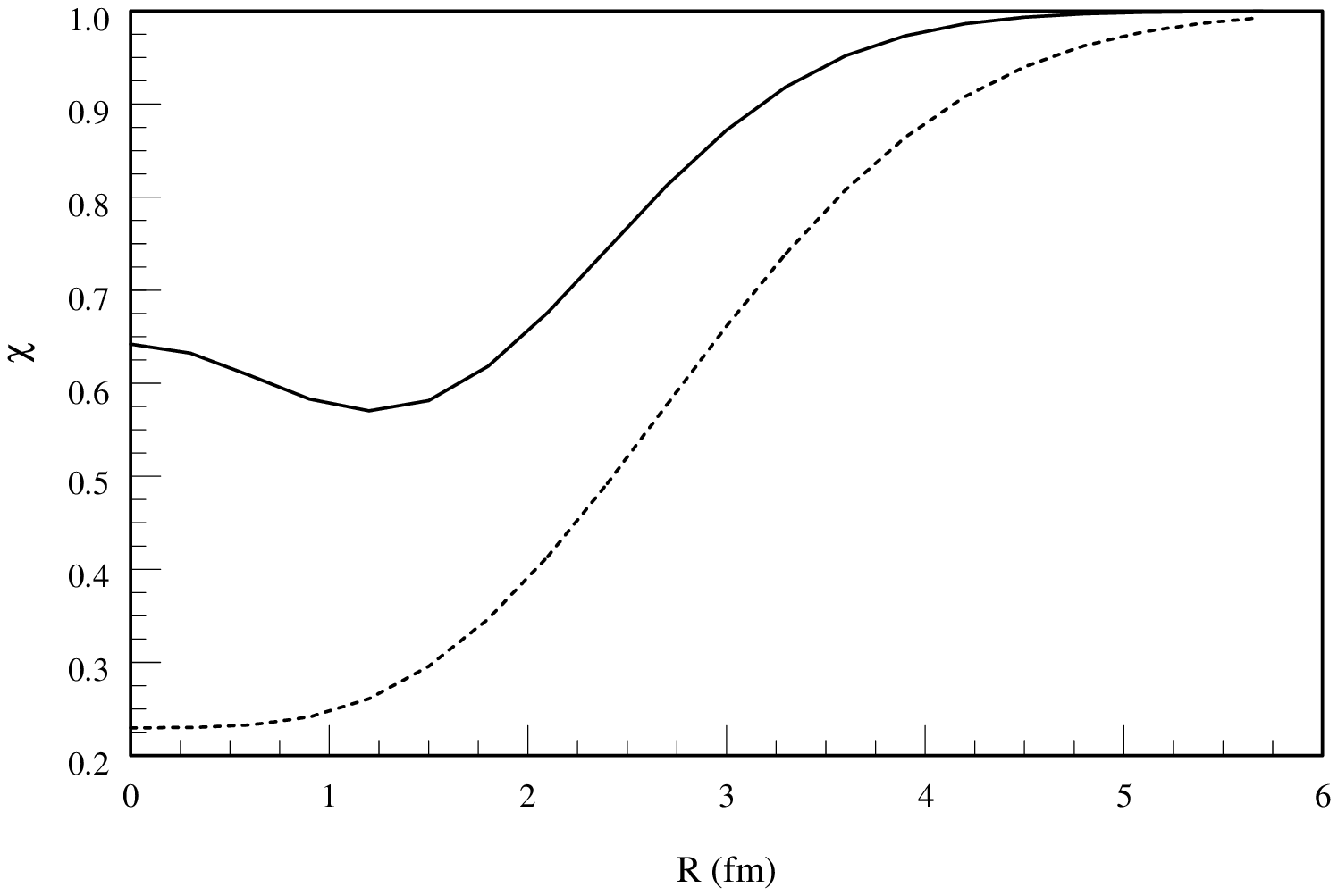}
\vsize=5 cm
\caption{\sl{ $\chi$ as a function of $R$ in $O^{16}$
for constant (modified) meson
fields inside the skyrmion, full line (dashed line)}}
\label{fig4}
\end{figure}

Another dramatic effect shows up in the isoscalar
root mean square radius of the nucleon. Figure 5 depicts
this radius as a function of the location of the center of the
skyrmion in $O^{16}$.
\begin{figure}[tb]
\epsffile{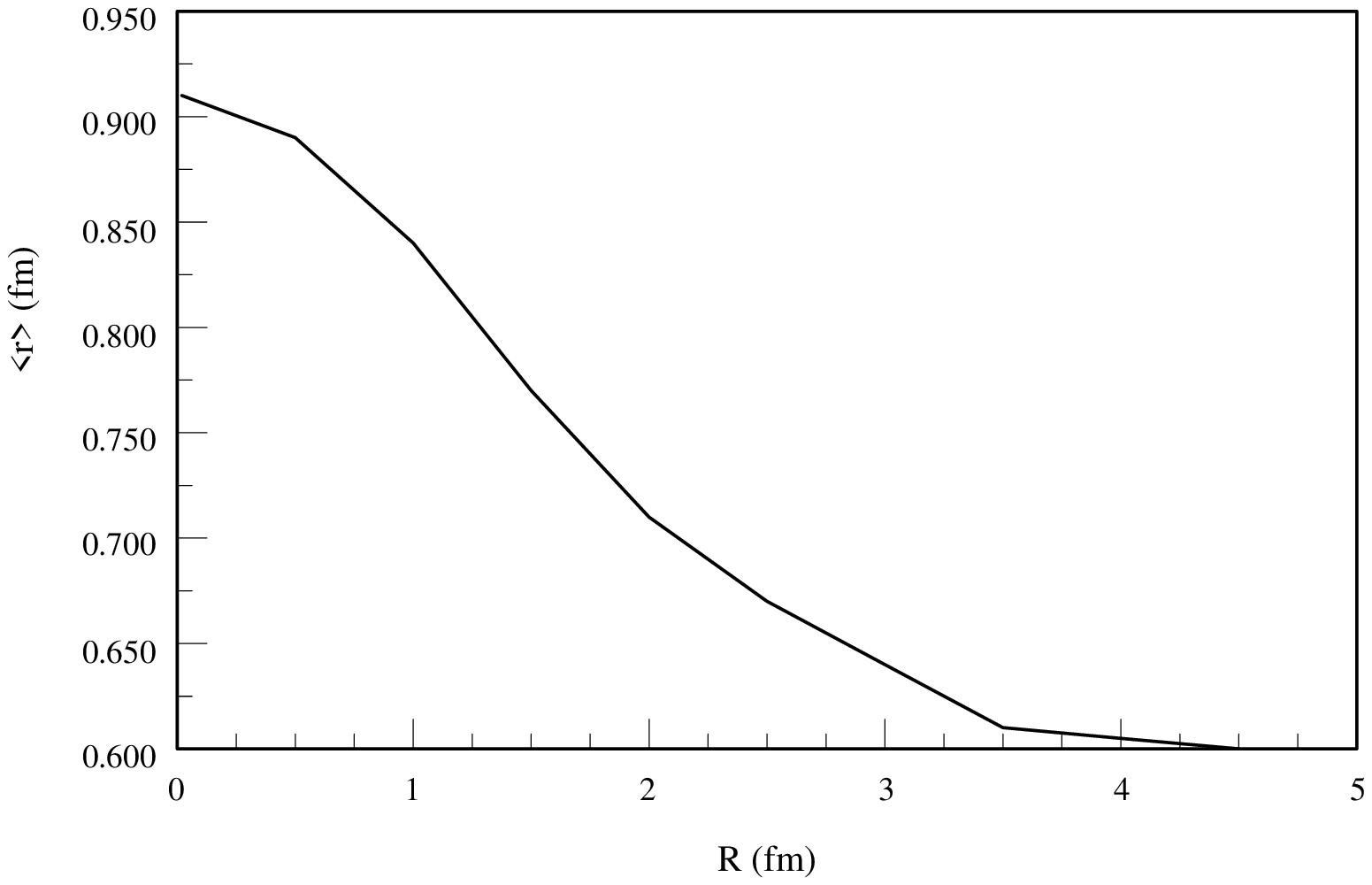}
\vsize=5 cm
\caption{\sl{ Skyrmion rms radius in $O^{16}$ with modified meson fields}}
\end{figure}
The nucleon swells inside the nucleus to almost twice its free
size. This is a bit troubling, because it would imply
a breakdown of the fluid approximation. The nucleons
no longer act as free molecules and the solitons overlap.
However, the overall picture does not change because
the strength of the interaction between solitons dimishes accordingly.
A measure of that interaction is the $\pi$NN coupling constant\ci{anw}.
A better estimate of the importance of the overlap
between skyrmions will demand a calculation of the
skyrmion-skyrmion interaction, but we limit ourselves
to obtain a rough estimate of the effect.

Figure 6 shows this constant as a function of distance for $O^{16}$.
It has the very opposite behavior. It drops to almost
half its value inside the nucleus. So, perhaps there is
still room to consider the nucleons as essentially free
objects. Despite the fact that they overlap, the
strength of the mutual interaction renders this overlap
quite ineffective.
\begin{figure}[tb]
\epsffile{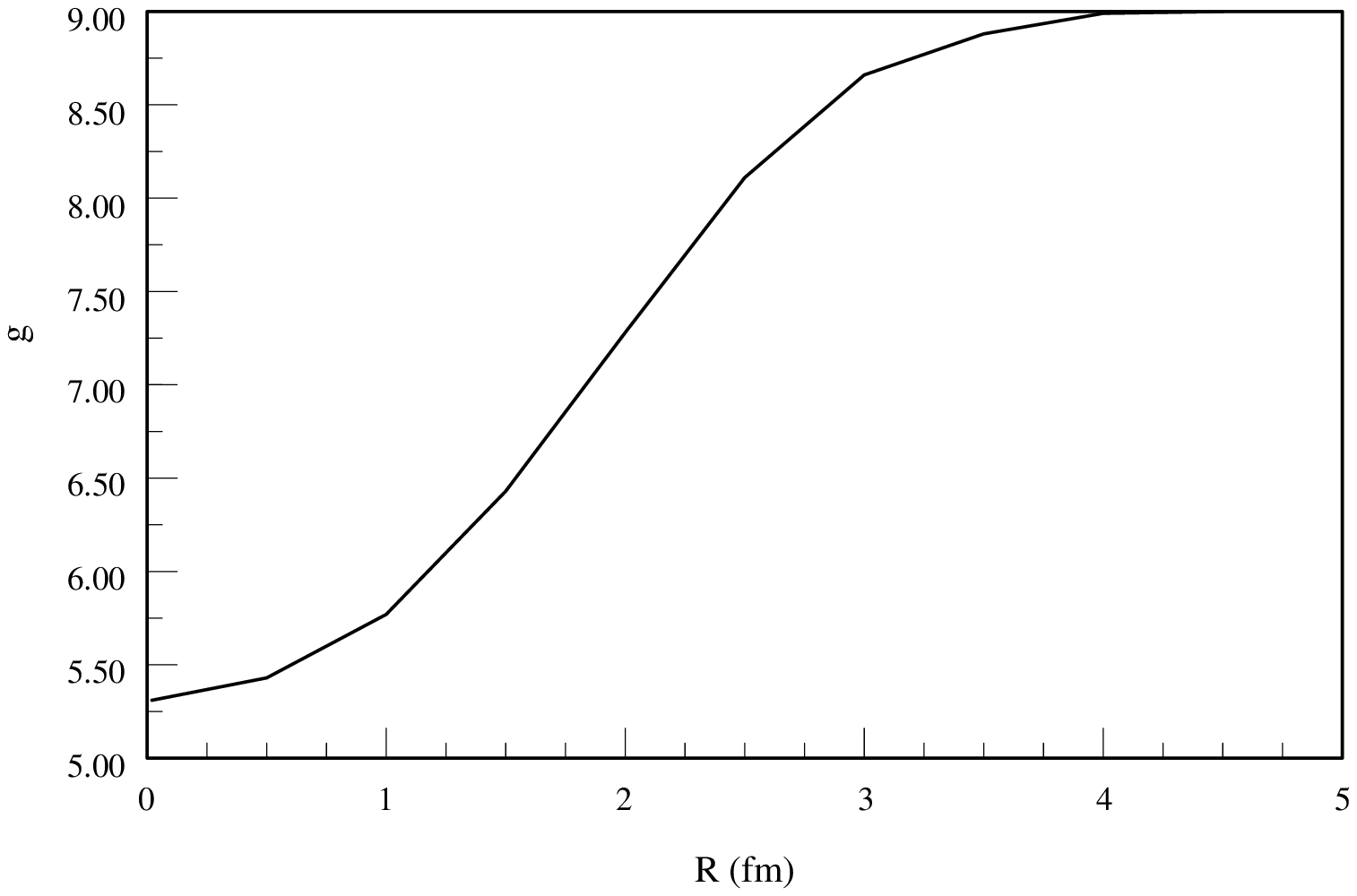}
\vsize=5 cm
\caption{\sl{ $g_{{\pi}NN}$ in $O^{16}$ with modified meson fields}}
\label{fig6}
\end{figure}

Other nucleon observables, such as magnetic g-factors,
 axial coupling constant and magnetic radii are less
susceptible to changes in the environment.

We have attempted to put the scalar fields we found in correspondence
with the Kisslinger type of potential with the identification\ci{koltun},
\ci{ericson}.

\bea\label{kisl}
\chi=\frac{-\kappa}{1+g'_0 \kappa}\nono
\kappa=4\pi~c_0\rho(R)
\eea

where $\rho$ is the nuclear density, $c_0\approx .21~m_{\pi}^{-3}$,
and $g'_0 < 1$. This procedure would predict the values of these
parameters as suggested by the Skyrme model.
Although it was possible to fit the scalar field with the above potential
around the surface of the nucleus, it is impossible to obtain
a moderately reasonable fit inside the nucleus.
In all cases there seems to be convergence to a set of
parameters that indicate a much larger Landau parameter than expected,
approximately around  $g'_0\approx 1$.
The Skyrme model appears to be suggesting a strong medium correction to
the so-called effective diffraction index. Inside the nucleus
nonlinearities play a dominant role.

\newpage

{\bf Acknowledgements}

This work was supported in part by the Department of
Energy under grant DE-FG03-93ER40773 and by the National Science Foundation
under grant PHY-9413872.

\end{document}